\documentclass[aps,pra,twocolumn,showpacs,floatfix]{revtex4}
\usepackage{epsfig}
\usepackage{graphicx}
\usepackage{dcolumn}
\usepackage{amsthm,amsmath}

\begin{document}

\title{Dispersion coefficients for the interactions of the alkali and alkaline-earth ions and inert gas atoms with a graphene layer}

\author{Kiranpreet Kaur and Bindiya Arora\footnote{Email: arorabindiya@gmail.com}}
\affiliation{Department of Physics, Guru Nanak Dev University, Amritsar, Punjab-143005, India}
\author{B. K. Sahoo\footnote{Email: bijaya@prl.res.in}}
\affiliation{Theoretical Physics Division, Physical Research Laboratory, Navrangpura, Ahmedabad-380009, India}
\date{Recieved Date; Accepted Date}

\begin{abstract}
Largely motivated by a number of applications, the van der Waals dispersion coefficients ($C_3$s) of the alkali ions (Li$^+$, 
Na$^+$, K$^+$ and Rb$^+$), the alkaline-earth ions (Ca$^+$, Sr$^+$, Ba$^+$ and Ra$^+$) and the inert gas atoms (He, Ne, Ar and Kr)
with a graphene layer are determined precisely within the framework of Dirac model. For these calculations, we have evaluated the dynamic 
polarizabilities of the above atomic systems very accurately by evaluating the transition matrix elements employing relativistic 
many-body methods and using the experimental values of the excitation energies. The dispersion coefficients are, finally, given as 
functions of the separation distance of an atomic system from the graphene layer and the ambiance temperature during the interactions. 
For easy extraction of these coefficients, we give a logistic fit to the functional forms of the dispersion coefficients in terms of 
the separation distances at the room temperature.
\end{abstract}
\maketitle

\section{Introduction}\label{sec1}
Since the carbon nanostructures are highly sensitive to their thermal, mechanical and electrical properties, they are 
extensively used both for the scientific and industrial applications. And hence, their studies are of great importance in 
the scientific community ~\cite{novoselov1,novoselov2,friedrich}. Some of the prominent applications include their utility
in nanotechnology, biochemical sensors, optics, electronics, new composite materials ~\cite{iijima,ebbesen,guo}, ion storage, 
nano electromechanical systems (NEMS) and ion channeling in carbon nanotubes (CNTs), secured wireless connections, efficient 
communication devices etc. \cite{mow,novo}. Among various carbon nanostructures, graphene, a one atom thick layer of carbon with remarkable 
properties, has been recently given considerable attention. They are of much significance in the areas of development of sensor technologies 
\cite{zhang,novo}, encapsulation of drugs \cite{chan,hill,novo}, nanofiltration membranes, regulating carbon dioxide for tackling climate 
change etc.  
Moreover, it has been observed that interaction of graphene with various species like atoms, molecules or ions can change its electronic and magnetic properties 
~\cite{tang,wang} and is being studied extensively in
context to the phenomenon of quantum-reflection. Interactions of the alkali metal atoms with a graphene layer have been recently investigated in Ref.~\cite{kiran} and 
with a single walled CNT were analyzed in Ref.~\cite{harjeet}. Since these interactions are extremely weak, it is immensely
difficult to measure them precisely using any experimental technique. Instead, sophisticated theoretical studies are carried out to find 
them more reliably. A number of calculations are reported using a wide variety of many-body methods, such as density functional theories 
~\cite{dino,dft1,dft2,dft3,dft4}, lower order many-body methods ~\cite{mbpt1}, Lifshitz approximations~\cite{dirac-hydro,
caride,som1,babb,dirac-hydro2} etc, to study the nature of interaction of the carbon nanostructures with various materials like 
ions, atoms, molecules etc. For example, Klimchitskaya and co-workers have explained the interaction of graphene layer with metal 
plates~\cite{bordag,jiang,winter} and atomic systems such as H ~\cite{dirac-hydro}, Na, Rb, Cs ~\cite{dirac-hydro,dirac-hydro2}, 
H$_2$ molecule ~\cite{dirac-hydro}, He$^+$ ion ~\cite{dirac-hydro,dirac-hydro2} etc using Lifshitz theory. Due to the
significance of studying atomic system and graphene interactions accurately, it would be useful to explore
behavior of these interactions for other atomic systems such as the presently considered alkali ions, alkaline-earth ions, 
and inert gas atoms with a graphene layer. Primary interests of choosing these particular atomic systems are for their applications to the modern technology.  For instance, the interaction of the lithium ion (Li$^+$) with graphene has applications in enhancing lithium storage capacity in 
lithium ion cells \cite{dress,yang} and improving performance of rechargeable lithium ion batteries ~\cite{novo,konga}. Similarly, the 
interaction of the alkaline earth ions with the carbon nanostructures have potential applications in the heterogeneous 
catalysis, bio-sensing \cite{novo}, hydrogen storage ~\cite{intro3,intro4,intro5} for powering green vehicles, molecular 
seiving, water desalination etc.. In the Lifshitz theory, these interactions can be explained using two models: 
hydrodynamic model ~\cite{bordag3,bordag5,blagov} and Dirac model ~\cite{bordag2}. Among these two models, Dirac model is
more adequate ~\cite{churkin} since it considers the quasi-particle fermion excitations in the graphene as massless Dirac fermions moving 
with the fermi velocity. Accuracies in the determination of the atom-wall interactions also depend on the accuracies of the dynamic 
polarizabilities of the atomic systems that appear in the formulae of the Lifshitz theory. For instance, the roles of using accurate values 
of the dynamic polarizabilities of the alkali atoms to describe interactions of these atoms with a graphene layer both in the hydrodynamic 
and Dirac models at zero temperature have been emphasized in Ref.~\cite{harjeet}. In this work, we intend to calculate the dispersion 
coefficients of the alkali ions (Li$^+$, Na$^+$, K$^+$ and, Rb$^+$), alkaline-earth ions (Ca$^+$, Sr$^+$, Ba$^+$ and, Ra$^+$), and inert 
gas atoms (He, Ne, Ar, Kr, and Xe) with a graphene layer at room temperature using accurately estimated polarizability values.

\section{Theory of Dispersion Coefficient}\label{sec2}

The general expression of van der Waals and Casimir Polder energy in terms of the dispersion $C_3$ coefficient for an atomic
system interacting with a graphene layer is expressed as ~\cite{dirac-hydro}
\begin{eqnarray}
E(a)= - \frac{C_3}{a^3},
\end{eqnarray}
where $a$ is the separation distance between the atom or ion from the graphene layer. The explicit expressions for the $C_3$ coefficients at 
zero temperature and non-zero temperature (in Kelvin), in terms of the reflection coefficients ${r_{TM}}$ and ${r_{TE}}$, 
are given by~\cite{advbook,dirac-hydro2,kiran}
\begin{eqnarray}
C_3(a)&=&-\frac{1}{16\pi}\int_0^{\infty}d\xi\alpha(\iota\xi)\int_{2a\xi \alpha_{fs} }^{\infty}dye^{-y}y^2 \nonumber\\
 & &\left(2r_{{TM}}-\frac{4a^2 \alpha_{fs}^2\xi^2}{y^2}(r_{{TM}}+r_{{TE}})\right)
\end{eqnarray}
and
\begin{eqnarray}
C_3(a,T)&=&- {\frac{k_B T}{8}} {\sum'_l} \alpha(\iota \zeta_l \omega_c) \int_{\zeta_l}^{\infty}dy \{ e^{-y} 2y^2 {\zeta_l}^2 \nonumber \\
&& r_{\rm{TM}}(\iota\zeta_l,y) \left[r_{\rm{TM}}(\iota\zeta_l,y)+r_{\rm{TE}}(\iota\zeta_l,y)\right]\},\label{eq-u}
\end{eqnarray}
respectively, where $\alpha_{fs}$ is the fine structure constant and $\alpha(\iota\omega)$ is the dynamic polarizability of the respective 
atomic system along imaginary frequency $\iota \omega$. In the above expressions, it is assumed that graphene is in thermal equilibrium at temperature $T$. 
It is obvious from the above expressions that accurate estimate of $C_3$ coefficients require accurate values of the  dynamic 
polarizabilities $\alpha(\iota{\xi_l})$ along the imaginary matsubara frequencies, ${\xi_l}=2 \pi k_B T l/{\hbar}$ with $l=0, 1, 2,..,$ 
of the considered atomic systems. It should also be noted that the prime over the summation sign in the above expression indicates 
multiplication of $l=0$ term with a factor of $1/2$. The reflection coefficients of the electromagnetic oscillations on graphene 
in the Dirac model defined at nonzero temperature are given by ~\cite{semen,dvince,dft3,drosdoff,dr1,dr2,dr3}
\begin{eqnarray}
r_{\rm{TM}}(\iota\zeta_l,y)&=&\frac{y{\tilde{\Pi}}_{00}}{y{\tilde{\Pi}}_{00}+2(y^2-\zeta_l^2)}
\end{eqnarray}
and
\begin{eqnarray}
r_{\rm{TE}}(\iota\zeta_l,y)&=&-\frac{(y^2-\zeta_l^2)\tilde{\Pi}_{tr}-y^2\tilde{\Pi}_{00}}{(y^2-\zeta_l^2)(\tilde{\Pi}_{tr}+2y)-y^2\tilde{\Pi}_{00}},~\label{eq-dirac}
\end{eqnarray} 
where $\tilde{\Pi}_{00}$ and $\tilde{\Pi}_{\rm tr}$ are the components of dimensionless polarization tensors given 
in \cite{dirac-hydro2,ivf}. The above expressions include a certain physical quantity $\Delta$, known as the gap parameter.
This parameter is accustomed to instigate the atom-graphene interaction coefficient. Although the exact value of $\Delta$ is
unknown, its maximum value is assumed to be 0.1 eV. However, we take $\Delta=0.01$ eV throughout the paper.

In the present work, we take into account the reflection coefficients at zero and non-zero temperatures from the previous studies 
~\cite{bordag2,dirac-hydro2,ivf}. We give more emphasis here on the use of precise values of the dynamic polarizabilities in the 
determination of the $C_3$ coefficients in the interactions of the considered atomic systems with a graphene layer. In the 
following section, we discuss briefly about the approaches adopted to evaluate these polarizabilities.

\section{Approaches to Evaluate Polarizabilities}

 The expression for the dynamic dipole polarizability of an atomic state $|\Psi_0^{(0)} \rangle$ with an imaginary frequency $\iota \omega$ 
is given by
\begin{equation}
\alpha (\iota \omega) =- 2 \frac{\langle \Psi_0^{(0)}|D|\Psi_0^{(1)} \rangle}{ \langle \Psi_0^{(0)}| \Psi_0^{(0)} \rangle },
\label{eq3}
\end{equation}
where $|\Psi_0^{(1)} \rangle$ is the first-order perturbed wave function to $|\Psi_0^{(0)} \rangle$ due to the dipole 
operator $D$ and is the solution of the first order differential equation
\begin{eqnarray}
(H-E_0^{(0)}-\iota \omega) |\Psi_0^{(1)} \rangle &=& \frac{(E_0-H)D}{H-E_0+ \iota \omega}|\Psi_0^{(0)} \rangle,
\label{eq4}
\end{eqnarray}
for the atomic Hamiltonian $H$, which is taken in the Dirac-Coulomb approximation for the present work, and $E_0^{(0)}$ is the energy 
eigenvalue corresponding to the state $|\Psi_0^{(0)} \rangle$. In the above expression, difficulties with the accurate estimate of 
$\alpha$s lie in the determination of both $|\Psi_0^{(0)} \rangle$ and $|\Psi_0^{(1)} \rangle$ of an atomic system. One can also write the 
above expression in the sum-over-states approach as
\begin{equation}
\alpha(\iota \omega)=-\frac{2}{\langle \Psi_0^{(0)}| \Psi_0^{(0)} \rangle } \sum_{I\ne 0}\frac{(E_0^{(0)}-E_I^{(0)})|\langle \Psi_0^{(0)}|D|\Psi_I^{(0)} \rangle|^2}{E_0^{(0)}-E_I^{(0)}+\omega^2},
\label{eq2}
\end{equation}
where $I$ represents all possible allowed intermediate states with their corresponding energies $E_I^{(0)}$s. This approach 
can be conveniently employed to the one-valence atomic systems like the alkali atoms and singly charged alkaline-earth metal 
ions to determine their polarizabilities as the matrix elements $\langle \Psi_0^{(0)}|D|\Psi_I^{(0)} \rangle$ among a large 
intermediate states of these systems can be calculated using the Fock-space relativistic coupled-cluster (RCC) method as have been 
demonstrated elaborately in our previous works \cite{jasmeet} and the excitation energies can be taken from the measurements. We use the 
polarizabilties of the alkaline-earth ions that were given in our previous work \cite{jasmeet}, but the polarizabilities for the alkali 
ions and inert noble gas atoms are obtained using the following procedure. 
\begin{table}[t]
\caption{\label{pol} Comparison of the scalar polarizabilities ($\alpha(0)$s) of the alkali ions (Li$^+$, Na$^+$, K$^+$, Rb$^+$), alkaline 
earth ions (Ca$^+$, Sr$^+$, Ba$^+$, Ra$^+$) and inert gas atoms (He, Ne, Ar, Kr) from different theoretical and experimental works. 
References are given inside the square brackets.}
\begin{ruledtabular}
\begin{tabular}{llll}
{System}  & This work & Others & Experiment\\
 \hline
 & & \\
 Li$^+$  & 0.19 & 0.1894\cite{kolb}, 0.192486\cite{bhatia} & 0.188\cite{cooke}\\
     &      & 0.1913 \cite{yashpal1} & \\ 
 Na$^+$  & 0.95 & 0.9457\cite{kolb}, 1.00\cite{lim1} & 0.978\cite{opik}\\
      &      & 0.9984 \cite{yashpal1} & \\
 K$^+$   & 5.45 & 5.457\cite{kolb}, 5.52\cite{lim1} & 5.47\cite{opik}\\
      &      & 5.522 \cite{yashpal1} & \\
 Rb$^+$  & 9.06 & 9.076\cite{kolb}, 9.11\cite{lim1} & 9.0\cite{john}\\
      &      & 9.213 \cite{yashpal1} & \\
  \hline
 & & \\
 Ca$^+$  & 76.77 &  75.88\cite{lim}, 75.49\cite{zhang2} & 75.3\cite{edward}\\
         &       & 73.0 \cite{bijaya1}, 76.1\cite{bindiya2}, 75.5\cite{patil} & \\
 Sr$^+$  & 92.24 & 88.29 \cite{bijaya2}, 91.10\cite{lim} & 93.3\cite{bark}\\
         &        & 91.3\cite{bindiya4}, 91.47\cite{patil} & \\
 Ba$^+$  & 124.40 & 124.26 \cite{bijaya2}, 123.07\cite{lim} & 123.88\cite{snow}\\
         &        & 124.7\cite{patil} & \\
 Ra$^+$  & 105.91 & 105.37\cite{lim}, 106.5\cite{saf1} & \\
         &        & 104.54 \cite{bijaya2}, 106.12 \cite{bijaya3} & \\
         &        & 106.22\cite{rupsi} & \\
 \hline
 & & \\
 He  & 1.32 & 1.32\cite{kolb}, 1.383\cite{soldan} & 1.3838\cite{langoff}\\
     &      & 1.360 \cite{yashpal1} & \\
 Ne  & 2.37 & 2.38\cite{kolb}, 2.697\cite{naka} & 2.668\cite{langoff}\\
     &      & 2.652 \cite{yashpal1} & \\ 
 Ar  & 10.77 & 10.77\cite{kolb}, 11.22\cite{naka} & 11.091\cite{langoff}\\
     &      & 11.089 \cite{yashpal1} & \\  
 Kr  & 16.47 & 16.47\cite{kolb}, 16.8\cite{naka} & 16.74\cite{langoff}\\
     &       & 16.93 \cite{yashpal1} & \\
 \end{tabular} 
\end{ruledtabular}
\end{table}

It is not advisable to employ the sum-over-states approach to determine the polarizabilities of the atomic systems having inert gas 
atomic configurations as evaluation of the dipole (E1) matrix elements of the dipole operator among different intermediate states 
of these systems are extremely difficult and might require to employ an approach similar to the equation-of-motion based many-body 
theory for their evaluation. This will demand large computational resources and sometime it may not be possible to calculate the E1 matrix 
elements for a sufficiently large number of intermediate states to estimate the polarizabilities within the required accuracies. One of the 
other appropriate approaches to determine polarizabilities of these inert gas atomic systems within the RCC method framework are demonstrated 
in \cite{yashpal1,yashpal2,bijaya}. Use of these RCC methods is also time consuming and can demand large computational 
resources. Since the addressed problem requires dynamic polarizabilities for a large set of imaginary frequencies, employing the above 
mentioned RCC method is impractical within a stimulated time frame to analyze the dispersion coefficients for all the considered 
inert gas atoms. Moreover, the above methods are appropriate only for calculating scalar polarizabilities and dynamic polarizabilities with 
real frequency arguments after a slight modification in the methodology (details are irrelevant to describe here). But it 
cannot be applied adequately to determine dynamic polarizabilities with imaginary frequency arguments. It has been 
demonstrated in the earlier studies \cite{yashpal1,yashpal2,bijaya} that scalar polarizabilities of the inert gas atomic systems evaluated 
using the relativistic random phase-approximation (RPA) match reasonably well with their experimental values. Thus, consideration 
of RPA can be good enough to determine dynamic polarizabities of the inert gas atomic systems. Advantage of applying this method is twofolds: 
firstly, calculation of polarizability for a given frequency can be performed within a reasonable time frame and secondly, a slightly modified RPA can be 
employed to determine dynamic polarizabilities at the imaginary frequencies as demonstrated below.

In RPA, expression for the dipole polarizability is given by
\begin{eqnarray}
 \alpha (\iota \omega)  &=& 2 \langle \Phi_0| D |\Psi_{RPA}^{(1)} \rangle.
\end{eqnarray}
This clearly suggests that wave function $|\Psi_0^{(0)}\rangle$ in Eq. (\ref{eq3}) is approximated to $|\Phi_0\rangle$, which 
is nothing but a mean-field wave function and is obtained using the Dirac-Fock (DF) method in this work, and the first order 
perturbed wave function is given by $|\Psi_{RPA}^{(1)} \rangle$. In RPA framework, we obtain $|\Psi_{RPA}^{(1)} \rangle$ as
\begin{eqnarray}
| \Psi_{RPA}^{(1)} \rangle &=&  \sum_{\beta}^{\infty} \sum_{p,a} \Omega_{a \rightarrow p}^{(\beta, 1)} |\Phi_0\rangle \nonumber \\
    &=& \sum_{\beta=1}^{\infty} \sum_{pq,ab} { \{} \frac{[\langle pb | \frac{1}{r_{12}} | aq \rangle 
- \langle pb | \frac{1}{r_{12}} | qa \rangle] \Omega_{b \rightarrow q}^{(\beta-1,1)} } {(\epsilon_p - \epsilon_a)^2+\omega^2}  \nonumber \\ 
&& + \frac{ \Omega_{b \rightarrow q}^{{(\beta-1,1)}^{\dagger}}[\langle pq | \frac{1}{r_{12}} | ab \rangle - \langle pq | \frac{1}{r_{12}} | ba \rangle] 
}{(\epsilon_p-\epsilon_a)^2+\omega^2} { \}} \nonumber \\
 && \times (\epsilon_p -\epsilon_a) |\Phi_0\rangle,
\label{eqrpa}
\end{eqnarray} 
where $\Omega_{a \rightarrow p}^{(\beta, 1)}$ is a wave operator that excites an occupied orbital $a$ of $|\Phi_0 \rangle$ to a virtual 
orbital $p$ which alternatively refers to a singly excited state with respect to $|\Phi_0 \rangle$ with $\Omega_{a \rightarrow p}^{(0,1)} 
= \frac{ \langle p | (\epsilon_p-\epsilon_a) D | a \rangle} {(\epsilon_p - \epsilon_a)^2+\omega^2}$ for the single particle orbitals 
energies $\epsilon$s and the superscripts $\beta$  and 1 representing the number of the Coulomb ($\frac{1}{r_{12}}$ in atomic unit (au)) 
and $D$ operators, respectively. 

\begin{figure}[t]
\includegraphics[scale=0.67]{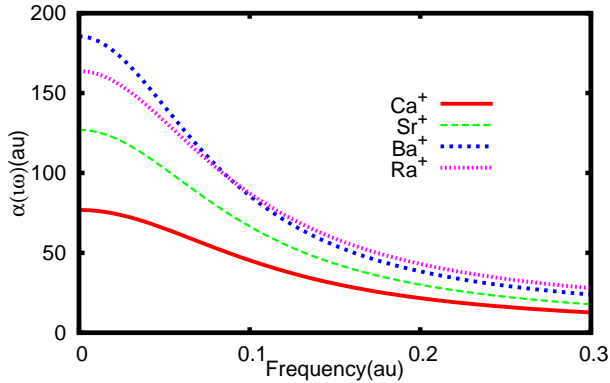}
\caption{(Color online) Dynamic polarizabilities of the alkaline earth ions Ca$^+$, Sr$^+$, Ba$^+$ and Ra$^+$ interacting with a graphene layer 
as functions of frequency.}~\label{pol1}
\end{figure}

\section{Results and Discussion}\label{sec7}

 We present the scalar polarizability ($\alpha(0)$) values for all the considered atomic systems in Table \ref{pol} obtained using our calculations 
and compare them against the results available from other theoretical studies using varieties of many-body methods and experimental
measurements. Among the other theoretical works, Johnson \textit{et al.} \cite{kolb} have performed the RPA calculations for the singly 
ionized alkali ions, our results are found to be consistent with their values. In another work, Soldan and co-workers \cite{soldan} have 
reported these values by employing coupled-cluster method. Lim \textit{et al.} \cite{lim1} have also evaluated these polarizabilities by employing
the RCC method and considering scalar relativistic atomic Hamiltonian, but their values are found to be larger than the RPA and 
experimental results. The reason could be that their approximated method may be overestimating the correlation effects beyond the RPA 
contributions. Nakajima and Hirao \cite{naka} have also investigated the polarizability values for inert gas systems using the 
relativistic effects in the estimate of $\alpha$ using the Douglas-Kroll (DK) Hamiltonian and adopting the finite gradient method.
Sahoo and co-workers report these values for many systems using the RCC method \cite{yashpal1,bijaya1,bijaya2}. The calculations by Patil 
\cite{patil} are carried out by using multipole matrix elements calculated from simple wave functions based on asymptotic behavior 
and on the binding energies of the valence electron. Safronova and co-workers \cite{saf1} have calculated the polarizabilities using
relativistic all-order single double method where all the single and double excitations of the Dirac-Fock wave function are included to all orders of 
perturbation theory.
For Li$^+$ ion, Cooke \textit{et al.} \cite{cooke} have determined the dipole polarizability from the \textit{d-f} and 
\textit{d-g} energy splittings using a laser excitation and optical detection scheme. The dipole polarizabilities of closed-shell
Na$^+$ and K$^+$ ions are obtained from observed spectra, using theoretical values of quadrupole polarizabilities taken from literature and including a 
number of corrections up to the fourth order in Ref.\cite{opik}. Our results for Rb$^+$ ion are in close agreement with the experimental values given 
by Johansson \cite{john}. Experimental analysis of the dipole polarizability values for the Ca$^+$ ion has been done by Chang \cite{edward}.
However, the ground-state polarizability of the Sr$^+$ ion given in \cite{bark} by Barklem and OMara using oscillator strength sum rules has a considerable
discrepancy with our results.
Snow and Lundeen \cite{snow} have performed high precision measurements for calculating polarizability of Ba$^+$ ion using a novel technique 
based on resonant Stark ionization spectroscopy microwave technique. We observe that our results are in agreement with the experimental values.
A noticeable variance in the polarizabilities of inert atoms Ne and Ar from the experimental results \cite{langoff} by Langhoff and Karplus is 
seen, in which they have employed a method based on Cauchy dispersion equation and Pad\'{e} approximates are used for extrapolation which improves the
convergence of the Cauchy equation.

\begin{figure}[t]
\includegraphics[scale=0.67]{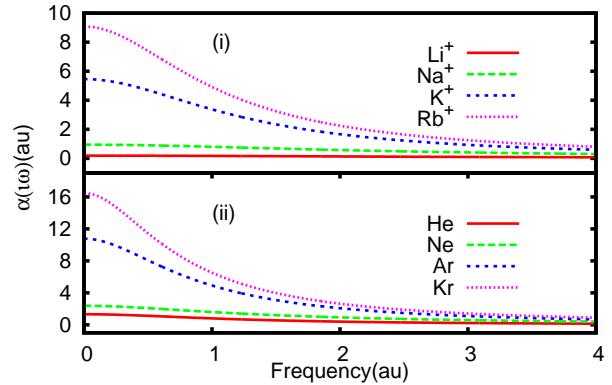}
\caption{(Color online) Dynamic polarizabilities of the (i) alkali ions Li$^+$, Na$^+$, K$^+$ and Rb$^+$, and (ii) inert gas atoms He, Ne, Ar and Kr 
interacting with a graphene layer as functions of frequency.}~\label{pol2}
\end{figure}
Comparison between these results show that our methods are giving reasonably accurate $\alpha(0)$ values, thus these methods can be employed 
to determine dynamic polarizabilities in these atomic systems within the similar accuracies as observed in the evaluation of the scalar 
polarizabilities. We plot these dynamic polarizabilities in Figs. \ref{pol1} and \ref{pol2}. As seen from the figures, alkaline earth ions 
have the highest polarizabilities, followed by inert gas atoms and then alkali ions.

 Using the dynamic polarizabilities given above, we now determine the dispersion coefficients of the considered alkali ions, alkaline-earth
ions, and inert gas atoms interacting with a graphene layer with the reflection coefficients estimated using the Dirac model. This is an 
extension of our previous work ~\cite{kiran}, where the interaction of the alkali atoms with a graphene layer is investigated. Here, 
we adopt the approaches described in Refs. ~\cite{harjeet,dirac-hydro2} to evaluate the reflection coefficients for the determination 
of $C_3$ coefficients as a function of separation distance of an atomic system from the graphene layer and a function of temperature. These 
results are discussed systematically below for each class of atomic systems.

\begin{figure}[t]
\includegraphics[scale=0.68]{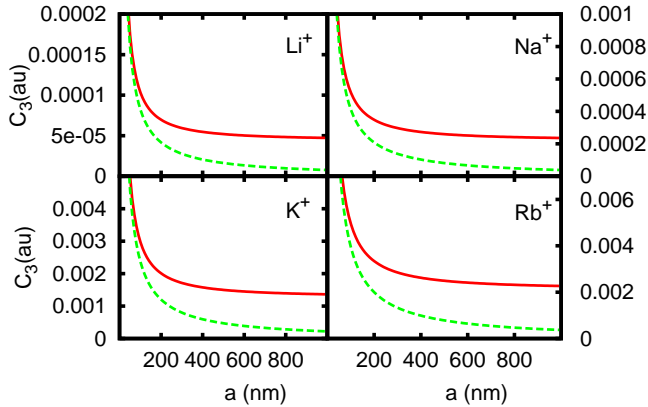}
\caption{(Color online) The $C_3$ coefficients (in au) as function of the ion-graphene separation distance for the 
alkali metal ions Li$^+$, Na$^+$, K$^+$ and Rb$^+$ interacting at $T=300^{\circ}$ K (solid red curve) and 
$T=0^{\circ}$ K (dashed green curve).}~\label{alkali}
\end{figure}

\subsection{Interactions of alkali ions with a graphene layer}

In Fig. \ref{alkali}, the graph between $C_3$ coefficients as a function of the separation distance $a$ (in nm) for the interactions of
alkali ions Li$^+$, Na$^+$, K$^+$ and Rb$^+$ with a graphene layer is shown for a gap parameter $\Delta=0.01$ eV. The solid red curve 
corresponds to the room temperature $T=300$ K while the dashed green curve represents $T=0$ K temperature $C_3$ coefficients. It should 
be noted that the reflection coefficients being same for a particular interacting surface (in our case, graphene), at a specific
separation distance and temperature, the $C_3$ coefficients of an element completely depends on its dynamic dipole polarizabilities. 
It was found in ~\cite{harjeet} that the $C_3$ coefficients increases with corresponding increase in atomic sizes 
of the alkali atoms for a given separation distance. In the same manner, it is seen that there is a likewise increase in the dispersion 
$C_3$ coefficients with the increase in the size of alkali ions at a particular distance of separation.
It can also be observed from this figure that the interaction between these ions and a graphene layer are more effective at the short separations while 
they are negligible at large separation distances.

\begin{figure}[t]
\includegraphics[scale=0.67]{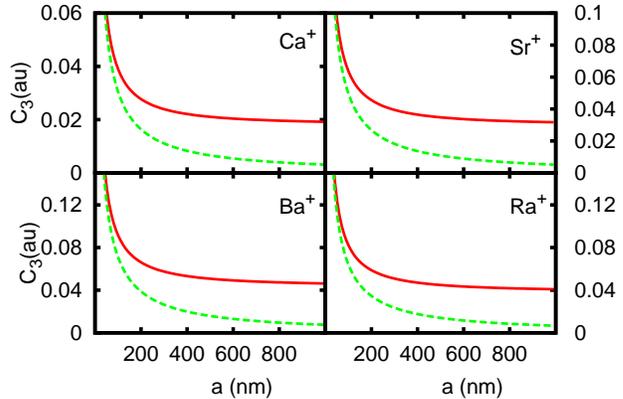}
\caption{(Color online) The $C_3$ coefficients (in au) calculated for the interactions of the alkaline-earth ions
Ca$^+$, Sr$^+$, Ba$^+$ and Ra$^+$ with a graphene layer as function of separation distance `$a$' (in nm) at temperatures 
$T=300^{\circ}$ K (red solid curve) and $T=0^{\circ}$ K (green dashed curve).}~\label{alkaline}
\end{figure}

\begin{figure}[t]
\includegraphics[scale=0.67]{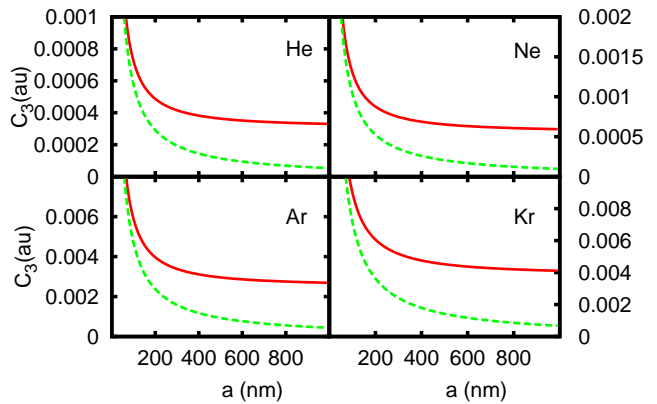}
\caption{(Color online) The $C_3$ coefficients (in au) for the interactions of the inert gas atoms He, Ne, Ar and Kr with graphene
as function of the separation distance `$a$' (in nm) at temperatures $T=300^{\circ}$ K (shown solid red curve) and $T=0^{\circ}$ K 
(shown dashed green curve).}~\label{inert}
\end{figure}

\subsection{Interactions of alkaline-earth ions with graphene}

The dispersion interactions of the alkaline-earth ions Ca$^+$, Sr$^+$, Ba$^+$ and Ra$^+$ with a graphene layer at temperatures $T=300$ K
(solid red curve) and $T=0$ K (dashed green curve) are shown in Fig. \ref{alkaline}.
It can be clearly seen from this figure that these $C_3$ coefficients are large for comparatively large ions except for the Ba$^+$ ion. 
This dominance of Ba$^+$ ion $C_3$ coefficient over the Ra$^+$ ion coefficient is due to the fact that the polarizabilty of Ba$^+$ ion is larger than
that of Ra$^+$ ~\cite{bijaya,saf1,jasmeet}. The reduction in the polarizability of Ra$^+$ ion is owing to the domineering contribution 
of the relativistic effects over the correlation effects ~\cite{lim}. Again, it can be seen in the figure that the interaction between 
these ions with a graphene layer are more effective at the short separations and  becomes insignificant at the large separation distances.
Among the three types of atomic systems, interaction of graphene with alkaline-earth ions is the strongest one. For a separation distance of 
300 nm, the interaction of alkaline earth ions with graphene layer is approximately 14 times stronger than the interaction with alkali 
ions, whereas approximately 8 times stronger than the interaction with inert gas atoms.

\subsection{Interactions of inert gas atoms with graphene}

The graph for the interactions between the inert gas atoms with a graphene layer, as a function of separation distance `$a$', is presented 
in Fig. \ref{inert}. We can clearly observe from the figure that $C_3$ coefficients of the ion-graphene interactions are large for 
comparatively large ions, owing to their greater values of scalar polarizabilities.
These coefficients are shown for temperatures $T=300$ K (solid red curve) and $T=0$ K (dashed green curve). It can be seen from Fig. 
\ref{inert} that the dispersion coefficients calculated at the room temperature show very less variation from the zero temperature 
coefficients at small separations whereas, at the larger distances of separations, we find a comparatively stronger dispersion interactions 
at $T=300$ K as compared to interaction at $T=0$ K.

\begin{table}[t]
\caption{\label{fit} Fitting parameters for the $C_3$(a, $T=300$ K) coefficients of the considered alkali ions, alkaline-earth ions and inert gas 
atoms with a graphene layer. A$_1$ and A$_2$ are given in the order of 10$^{-2}$ au and $x_0$ in nm.}
\begin{ruledtabular}
\begin{tabular}{ccccc}
 {Alkali ions}  & Li$^+$ & Na$^+$ & K$^+$ & Rb$^+$\\
 \hline
 & & \\
 A$_1$  & 0.48798 & 2.41443 & 13.3224 & 21.6418\\
 A$_2$  & 0.00368 & 0.01841 & 0.10702 & 0.17863\\
 $x_0$  & 1.5239 & 1.53425 & 1.58631 & 1.61405\\
  \hline
 {Alkaline earth ions}  & Ca$^+$ & Sr$^+$ & Ba$^+$ & Ra$^+$\\
 \hline
 & & \\
 A$_1$  & 66.8551 & 100.604 & 136.182 & 137.158\\
 A$_2$  & 1.71017 & 2.84881 & 4.19032 & 3.66618\\
 $x_0$  & 3.28993 & 3.49327 & 3.64421 & 3.31959\\
 \hline
 {Inert gas atoms}  & He & Ne & Ar & Kr\\
 \hline
 & & \\
 A$_1$  & 3.214 & 5.8524 & 24.7463 & 36.6345\\
 A$_2$  & 0.02594 & 0.04662 & 0.2133 & 0.32761\\
 $x_0$  & 1.59107 & 1.57743 & 1.65885 & 1.69824\\
 \end{tabular} 
\end{ruledtabular}
\end{table}

\subsection{Fitting Formula}

In contemplation of simplification in generating our results of $C_3$ coefficients for future theoretical and experimental verifications or
for extracting these values for various applications at room temperature with a given separation distance, we provide a logistic fit 
of the functional form of these coefficients as
\begin{equation}
 C_{3}(a)=A_2+\frac{A_1-A_2}{(1+{a/x_0})},
\end{equation}
where $A_1$ (in au), $A_2$ (in au) and $x_0$ (in nm)  are the fitting parameters that rely on the properties of the interacting atomic 
systems with a graphene layer. We give our fitting coefficients in Table ~\ref{fit} for extrapolating the dispersion coefficients for 
the considered elements-graphene layer interactions. We predict that obtained coefficients using above fitting parameters have
divergences not more than 6 \% with the coefficients calculated using Dirac model at $T=300$ K. Hence, the above equation serves as the
best suited fit to express the interactions of considered atomic systems with a graphene layer.

\section{Conclusion}

In summary, we have studied the dispersion interaction coefficients of the alkali-metal ions (Li$^+$, Na$^+$, K$^+$ and Rb$^+$), 
alkaline-earth ions (Ca$^+$, Sr$^+$, Ba$^+$ and Ra$^+$) and inert gas atoms (He, Ne, Ar and Kr) with a graphene layer. We have shown 
explicitly the dependence of these coefficients on the separation distance $a$ and temperature $T$.
We have commenced by using accurate values of the dynamic polarizabilities of the considered atomic systems by employing suitable 
relativistic many-body methods and calculating the reflection coefficients using the Dirac model. We observed that the dispersion 
interaction coefficients of the alkaline ions with the graphene layer is the strongest among the alkali ions and inert gas atoms  
interacting with graphene; the least dispersion interaction of graphene is with the alkali ions. It is also seen that due to the larger 
values of dynamic polarizabilities of the Ba$^+$ ion than that of Ra$^+$ ion, the dispersion coefficient of Ba$^+$ dominates over the 
Ra$^+$ ion. Our results can be of utmost use for the experimentalists in studying these interactions more reliably in view of the fact that 
performing these experiments at room temperature are comparatively more susceptible. This study also demonstrates about stronger dispersion
interactions of the alkaline ions, especially at the larger distances of separations, and  its consequences can be more applicably. In
addition, we also devise a promptly accessible functional form of logistic type having separation distance dependence at room temperature 
for easy extraction of the dispersion coefficients for the future applications.

\section*{Acknowledgement}
The work of B.A. is supported by CSIR grant no. 03(1268)/13/EMR-II, India. K.K. acknowledges the financial support 
from DST (letter no. DST/INSPIRE Fellowship/2013/758). B.K.S. acknowledges use of the PRL 3TFlop HPC cluster at Ahmedabad.


\end{document}